\documentclass[aps,pre,showpacs,amsmath,amssymb,longbibliography]{revtex4-1}
\usepackage{amsmath,amssymb,latexsym,epsfig,graphicx,epsf,bm}
\usepackage{graphicx}
\usepackage{color}
\usepackage{float}
\newcommand{\be}{\begin{equation}}
\newcommand{\ee}{\end{equation}}
\newcommand{\fig}[1]{Fig.~\ref{#1}}
\newcommand{\Fig}[1]{Figure~\ref{#1}}
\newcommand{\sect}[1]{Sec.~\ref{#1}}

\newcommand{\eq}[1]{Eq.~(\ref{#1})}
\newcommand{\Eq}[1]{Equation~(\ref{#1})}
\newcommand{\bet}{\bm{\eta}}
\newcommand{\Sex}{{S}_{\rm ex}}

\newcommand{\tSex}{\tilde{S}_{\rm ex}}
\newcommand{\Teq}{{T}_{\rm eq}}

\newcommand{\Tef}{{T}_{\rm eff}}
\newcommand{\Ts}{{T}_{\rm s}}
\newcommand{\Tg}{{T}_{\rm g}}
\newcommand{\Tf}{{T}_{\rm fic}}
\newcommand{\Tcon}{{T}_{\rm conf}}
\newcommand{\Tset}{{T}_{\rm s,1}}
\newcommand{\Tsto}{{T}_{\rm s,2}}

\newcommand{\br}{\bold r}

\newcommand{\bp}{\bold p}
\newcommand{\tbr}{\tilde{\bold r}}

\newcommand{\tbp}{\tilde{\bold p}}

\newcommand{\tm}{\tilde{m}}

\newcommand{\tnabla}{\tilde{\nabla}}
\newcommand{\txi}{\tilde{\xi}}
\newcommand{\bF}{\bold F}
\newcommand{\tbF}{\tilde{\bold F}}
\newcommand{\bR}{\bold R}
\newcommand{\tdg}{\dot{\tilde\gamma}}
\newcommand{\tbR}{\tilde{\bold R}}
\newcommand{\avm}{\langle m\rangle}
\newcommand{\bRa}{{\bf R}_{\rm a}}
\newcommand{\bRb}{{\bf R}_{\rm b}}
\renewcommand{\tt}{\tilde{t}}

\begin{document}
	\title{Isomorph theory beyond thermal equilibrium}
	\date{\today}
	\author{Jeppe C. Dyre}\email{dyre@ruc.dk}
	\affiliation{Glass and Time, IMFUFA, Department of Science and Environment, Roskilde University, P.O. Box 260, DK-4000 Roskilde, Denmark}	

\begin{abstract}
This paper generalizes isomorph theory to systems that are not in thermal equilibrium. The systems are assumed to be R-simple, i.e., have a potential energy that as a function of all particle coordinates $\bR$ obeys the hidden-scale-invariance condition $U(\bRa)<U(\bRb)\Rightarrow U(\lambda\bRa)<U(\lambda\bRb)$. ``Systemic isomorphs'' are introduced as lines of constant excess entropy in the phase diagram defined by density and systemic temperature, which is the temperature of the equilibrium state point with average potential energy equal to $U(\bR)$. The dynamics is invariant along a systemic isomorph if there is a constant ratio between the systemic and the bath temperature. In thermal equilibrium, the systemic temperature is equal to the bath temperature and the original isomorph formalism is recovered. The new approach rationalizes within a consistent framework previously published observations of isomorph invariance in simulations involving nonlinear steady-state shear flows, zero-temperature plastic flows, and glass-state isomorphs. The paper relates briefly to granular media, physical aging, and active matter. Finally, we discuss the possibility that the energy unit defining reduced quantities should be based on the systemic rather than the bath temperature.
\end{abstract}
\maketitle

\section{Introduction}

Isomorph theory explores the consequences of hidden scale invariance, which is the symmetry expressed by \eq{hsi} below \cite{sch14} in which $U(\bR)$ is the potential energy as a function of all $N$ particle coordinates $\bR\equiv (\br_1,...,\br_N)$ and $\bRa$ and $\bRb$ are two same-density configurations, 

\be\label{hsi}
U(\bRa)<U(\bRb)\,\,\Rightarrow\,\, U(\lambda\bRa)<U(\lambda\bRb)\,.
\ee
This logical implication states that the ordering of configurations at one density according to their potential energy is maintained if the configurations are scaled uniformly to a different density; for molecules the uniform scaling refers to the center of masses, keeping the spatial orientations and molecular sizes unchanged. 

Hidden scale invariance applies rigorously only for systems with an Euler-homogeneous potential-energy function (plus a constant). For realistic models, \eq{hsi} is fulfilled at best for most configurations and when the scaling parameter $\lambda$ is not far from unity. Nevertheless, \eq{hsi} and its consequences apply to a good approximation for the liquid and solid phases of many models. The thermodynamic phase diagram of a system with hidden scale invariance, an ``R-simple system'', is one-dimensional in regard to structure and dynamics. This is because there are lines in the phase diagram, the so-called isomorphs \cite{IV}, along which structure and dynamics in reduced units are invariant to a good approximation. Physically, isomorph invariance means that if one imagined filming the molecules, the same movie would be recorded at two different state points of a given isomorph except for a uniform scaling of space and time \cite{dyr18a}. 

\Eq{hsi} is referred to as ``hidden'' scale invariance because it is rarely obvious by inspection of the potential-energy function. For systems like the Lennard-Jones (LJ) system an explanation of \eq{hsi} is available in terms of an effective inverse-power-law pair potential plus a constant plus a linear term \cite{II}, but for instance for molecular systems we still do not know how to predict when \eq{hsi} applies to a good approximation. Fortunately, this is easily tested in simulations \cite{sch14,EXPII}, and one of the consequences of \eq{hsi} -- that of strong virial potential-energy correlations in the thermal-equilibrium fluctuations \cite{sch14} -- is also straightforward to check \cite{ped08,I,ing12b}.

The unit system defining reduced variables is based on the system's volume $V$ and temperature $T$. If the (number) density is $\rho\equiv N/V$, the length, energy, and time units are, respectively, \cite{IV}

\be\label{red_units}
l_0=\rho^{-1/3}\,\,,\,\,e_0=k_BT\,\,,\,\,t_0=\rho^{-1/3}\sqrt{\frac{\avm}{k_BT}}\,.
\ee
Here $\avm$ is the average particle mass. \Eq{red_units} refers to Newtonian dynamics; Brownian dynamics has the same length and energy units, but a different time unit \cite{IV}. All quantities can be made dimensionless by reference to the above units. The term ``reduced'' refers to the resulting dimensionless quantity. Reduced quantities are denoted by a tilde, for instance

\be\label{tbR}
\tbR
\,\equiv\,\rho^{1/3}\bR\,.
\ee

Although not widely used, the state-point-dependent unit system defined by \eq{red_units} is far from new. It was used already by Andrade in his papers on viscosity from the 1930s \cite{and30,and34} because this is the natural unit system when a liquid is approximated by a hard-sphere system \cite{dyr16}. Reduced units arise also in the proof that systems with an Euler-homogeneous potential-energy function have invariant physics along the lines in the thermodynamic phase diagram given by $\rho^{n/3}/T=$Const., in which the scaling exponent $n$ is defined by $U(\lambda\bR)=\lambda^{-n}U(\bR)$ \cite{kle19,hoo71,hiw74}. Finally, reduced units are crucial in Rosenfeld's seminal paper from 1977 introducing excess-entropy scaling \cite{ros77,dyr18a}. Reduced units are sometimes referred to as ``macroscopic'' because they are defined in terms of thermodynamic quantities \cite{hey15}, not in terms of microscopic quantities like the standard state-point-independent molecular-dynamics (MD) units \cite{tildesley,frenkel}.

The existence of isomorphs has mainly been validated in computer simulations, although some predictions of the isomorph theory have also been confirmed in experiments \cite{roe13,xia15,nis17,han18}. Computer simulated systems for which isomorph-theory predictions apply include, e.g., LJ-type liquids \cite{II,IV,V,yoo19}, the gas, liquid, and solid phases of the low-temperature EXP pair-potential system \cite{EXPI,EXPIV}, simple molecular models \cite{ing12b,fra19,kop20}, polydisperse systems \cite{ing15a}, crystals \cite{alb14}, nano-confined liquids \cite{ing13a}, polymer-like flexible molecules \cite{vel14}, metals \cite{hum15,fri19}, and plasmas \cite{vel15,tol19}. Density-scaling \cite{alb04,rol05,sch09,IV,gun11} and isochronal superposition \cite{rol03,nga05,nie08,roe13,han18} are examples of experimental findings that can be rationalized within the isomorph-theory framework, which incidentally also accounts for exceptions \cite{IV,dyr14,dyr18a} . Further regularities that may be explained by the isomorph theory include instantaneous equilibration for a jump on an isochrone \cite{IV,ing12b}, the quasiuniversality of simple liquids \cite{bac14a,dyr16}, how physical quantities vary along the melting line \cite{cos16,ped16}, excess-entropy scaling \cite{bel19,dyr18a}, and the Stokes-Einstein relation \cite{cos19}.  

The above examples all refer to equilibrium conditions, and indeed thermal equilibrium is a prerequisite of the current isomorph theory \cite{IV,sch14,dyr16}. A few papers have also demonstrated isomorph invariance under non-equilibrium conditions, however, such as steady-state shear flows of liquids and glasses \cite{sep13,ler14,jia19}. This shows the need for generalizing isomorph theory to systems that are not in thermal equilibrium, which is further emphasized by the fact demonstrated below that the explanations given so far for isomorph invariance in non-equilibrium systems are not consistent.

This paper shows that \eq{hsi} allows for a more general isomorph theory. Although the paper is self-contained, it will be easier to read for persons familiar with isomorph theory on the level of the reviews given in Refs. \onlinecite{ing12,dyr14,dyr16,dyr18a}.

\section{Background: The equilibrium theory}

This section summarizes the existing isomorph theory \cite{IV,sch14}. For a system in thermal equilibrium at (number) density $\rho$ and temperature $T$, the excess entropy $\Sex$ is defined as the entropy minus that of an ideal gas at the same density and temperature. Since an ideal gas is maximally disordered, one always has $\Sex\leq 0$. Any state point of the thermodynamic phase diagram is fully characterized by two thermodynamic variables, for instance: $\rho$ and $T$, $\rho$ and $\Sex$, $T$ and $\Sex$, $\rho$ and the average potential energy $U$, $U$ and $\Sex$, etc. We define the microscopic excess-entropy function $\Sex(\bR)$ by \cite{sch14}

\be\label{sex_def}
\Sex(\bR)
\,\equiv\, \Sex(\rho,U(\bR))\,.
\ee
This is short-hand notation for the following: if $\Sex(\rho,U)$ is the excess entropy of the equilibrium state point $(\rho,U)$, $\Sex(\bR)$ is defined as $\Sex(\rho,U)$ evaluated by substituting $U=U(\bR)$. It follows that, except for an additive constant, the microscopic excess entropy $\Sex(\bR)$ is the logarithm of the number of configurations with same density and potential energy as $\bR$. Note that $\Sex(\bR)$ is defined also if $\bR$ is not a typical equilibrium configuration of some state point. The only requirement is that the configuration is spatially homogeneous and, for instance, has no big holes, because otherwise a proper density cannot be identified. We shall henceforth only consider such configurations. Inverting \eq{sex_def} leads to

\be\label{any}
U(\bR)
\,=\,U(\rho,\Sex(\bR))\,
\ee
in which $U(\rho,\Sex)$ is the thermodynamic average potential energy of the state point $(\rho,\Sex)$. 

All said so far is general. Reference \onlinecite{sch14} showed that the hidden-scale-invariance condition \eq{hsi} implies that the function $\Sex(\bR)$ is scale invariant, i.e., $\Sex(\lambda\bR)=\Sex(\bR)$. In this case, $\Sex(\bR)$ depends merely on the configuration's reduced coordinate vector $\tbR$:

\be\label{Sex_rc}
\Sex(\bR)
\,=\,\Sex(\tbR)\,
\ee
and \eq{any} becomes

\be\label{fundeq}
U(\bR)
\,=\,U(\rho,\Sex(\tbR))\,.
\ee
This summarizes the 2014 version of isomorph theory \cite{sch14}, the original version of which appeared in 2009 \cite{IV}. 

All identities of the current isomorph theory may be derived from \eq{fundeq} \cite{sch14}, which is also the basis for the non-equilibrium generalization developed in the next section. For instance, \eq{fundeq} implies strong correlations between the constant-volume equilibrium fluctuations of virial $W$ and potential energy, $\Delta W\cong\gamma\Delta U$ \cite{ped08,I,II}, with the so-called density-scaling exponent $\gamma$ given \cite{IV} by

\be\label{gamma}
\gamma
\,\equiv\, \left(\frac{\partial\ln T}{\partial\ln \rho}\right)_{\Sex}
\,=\,\frac{\langle\Delta U \Delta W \rangle}{\langle(\Delta U)^2\rangle}\,.
\ee
The second equality sign is a general statistical-mechanical identity that allows for calculating $\gamma$ from constant-volume equilibrium fluctuations. If \eq{fundeq} were rigorously obeyed for all configurations, there would be perfect correlations, i.e., $\Delta W=\gamma\Delta U$, but as mentioned isomorph theory is usually only approximate.  

By means of the thermodynamic identity $T=(\partial U /\partial\Sex)_\rho$, a first-order Taylor expansion of \eq{fundeq} at the state point $(\rho,\Sex)$ leads \cite{sch14} to

\be\label{firstord}
U(\bR)\,\cong\, U(\rho,\Sex)\, +\, T(\rho,\Sex) \left(\Sex(\tbR)-\Sex\right)\,.
\ee
Consider now two equilibrium state points $(\rho_1,T_1)$ and $(\rho_2,T_2)$ with average potential energies $U_1$ and $U_2$ and the same excess entropy $\Sex$. Suppose $\bR_1$ and $\bR_2$ are equilibrium configurations of the state points with the same reduced coordinates, i.e., obeying $\rho_1^{1/3}\bR_1=\rho_2^{1/3}\bR_2\equiv \tbR$. By elimination of the common factor $\Sex(\tbR)-\Sex$, \eq{firstord} then implies that with $T_1\equiv T(\rho_1,\Sex)$ and $T_2\equiv T(\rho_2,\Sex)$ one has

\be\label{isom}
\frac{U(\bR_1)-U_1}{k_B T_1}\,\cong\,\frac{U(\bR_2)-U_2}{k_B T_2}\,.
\ee
This means that (in which $C_{12}$ is a constant)

\be\label{isomeq}
e^{-{U(\bR_1)}/{k_B T_1}}\,\cong\, C_{12}\,e^{-{U(\bR_2)}/{k_B T_2}}\,.
\ee
\Eq{isomeq} is the 2009 definition of an isomorph in the equilibrium phase diagram \cite{IV}, stating that along an isomorph the canonical probabilities of configurations that scale uniformly into one another are identical ($C_{12}$ disappears when the probabilities are normalized). It was assumed that the system in question is ``strongly correlating'' (=R-simple) in the sense that the equilibrium constant-density virial potential-energy fluctuations have a Pearson correlation coefficient larger than $0.9$. At the time, isomorphs were not defined to be configurational adiabats ($\Sex=$ Const.), but shown to be so from \eq{isomeq}. In contrast, the 2014 version of the theory \textit{defines} isomorphs as the configurational adiabats of an R-simple system \cite{sch14}.

\Eq{fundeq} implies invariant dynamics along isomorphs because the reduced force depends only on a given configuration's reduced coordinates. To demonstrate this we define the collective force vector $\bF$ as the vector of all particle forces $\bF\equiv (\bF_1, ..., \bF_N)$. It is straightforward to show that Newton's second law in reduced coordinates is $\tbF=d^2\tbR/d\tt^2$, assuming here for simplicity identical particle masses (absorbed into the reduced time). If the reduced force $\tbF$ depends only on a given configuration's reduced coordinates, the equation of motion has no reference to the density and is therefore the same for configurations that scale uniformly into one another, i.e., along an isomorph. This implies isomorph-invariant dynamics.

To show that $\tbF=\tbF(\tbR)$ for an equilibrium R-simple system, note that according to \eq{red_units} the reduced force is given by $\tbF=\rho^{-1/3}\bF/k_BT$ (a force times a length is an energy). Since $\bF=-\nabla U(\bR)$ we get $\tbF=-\rho^{-1/3}\nabla U(\bR)/k_BT$, which via $\rho^{-1/3}\nabla=\tilde{\nabla}$ and \eq{fundeq} implies that

\be\label{tbFeq}
\tbF
\,=\,-\tnabla U(\rho,\Sex(\tbR))/k_BT
\,=\,-\left(\frac{\partial U(\rho,\Sex(\tbR))}{\partial\Sex}\right)_\rho\tnabla\Sex(\tbR)/k_BT\,.
\ee
The notation $\left({\partial U(\rho,\Sex(\tbR))}/{\partial\Sex}\right)_\rho$ means the standard thermodynamic derivative $\left({\partial U(\rho,\Sex)}/{\partial\Sex}\right)_\rho$ into which $\Sex=\Sex(\tbR)$ is substituted. 
Recalling that $T=(\partial U /\partial\Sex)_\rho$, \eq{tbFeq} becomes in terms of the reduced excess entropy $\tSex\equiv\Sex/k_B$ 

\be\label{red_force}
\tbF
\,=\,-\tnabla\tSex(\tbR)\,.
\ee
This demonstrates that $\tbF$ for equilibrium configurations is a function only of the configurations' reduced coordinates, ensuring invariant dynamics along the isomorphs. If the reduced dynamics is isomorph invariant, by time averaging one finds as a consequence invariance of the reduced-unit structure. Thus both structure and dynamics are invariant along a systemic isomorph whenever the reduced force is a function of the reduced coordinates.

\begin{figure*}[htbp!]
	\begin{center}
		\includegraphics[width=80mm]{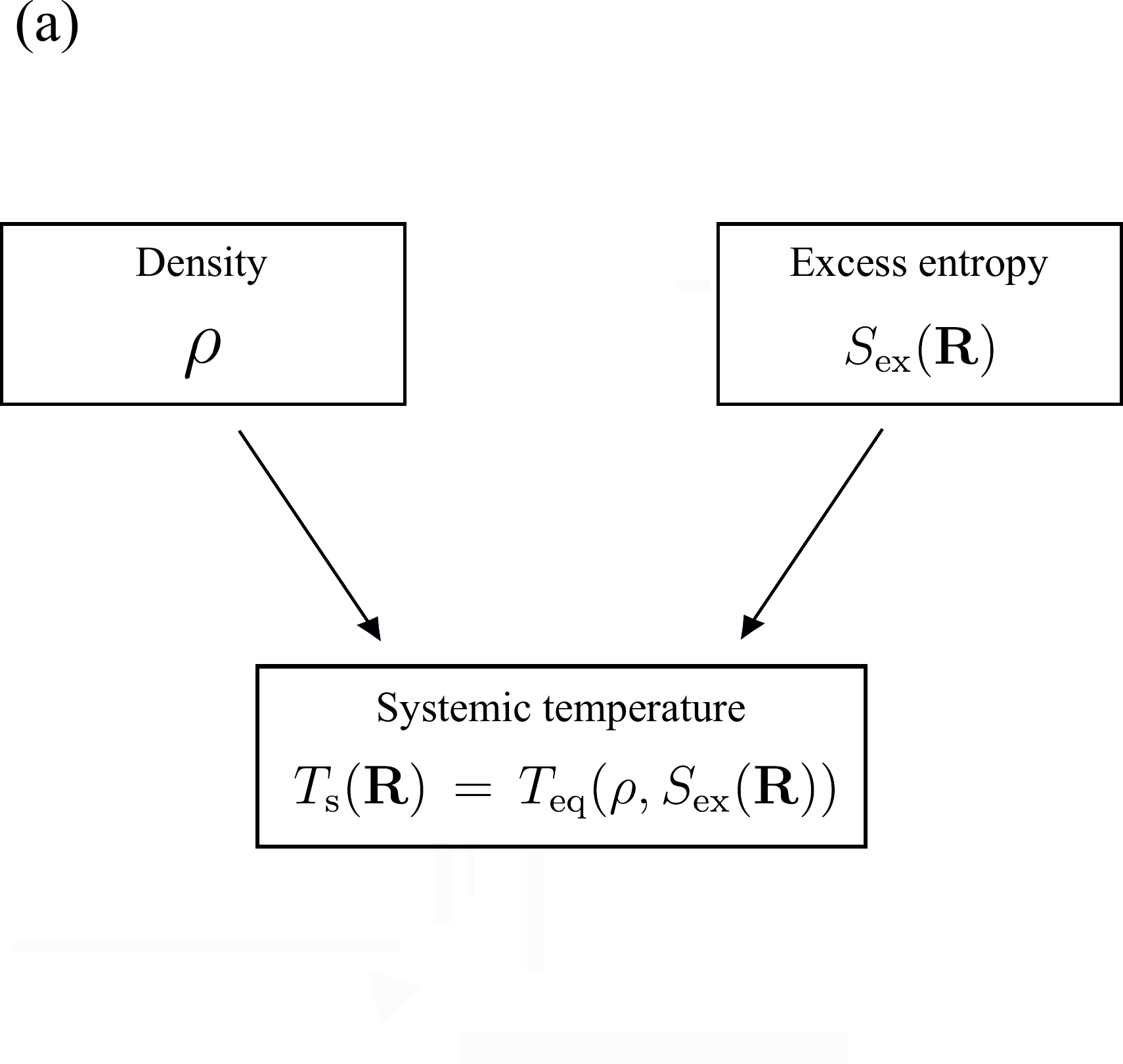}
		\includegraphics[width=80mm]{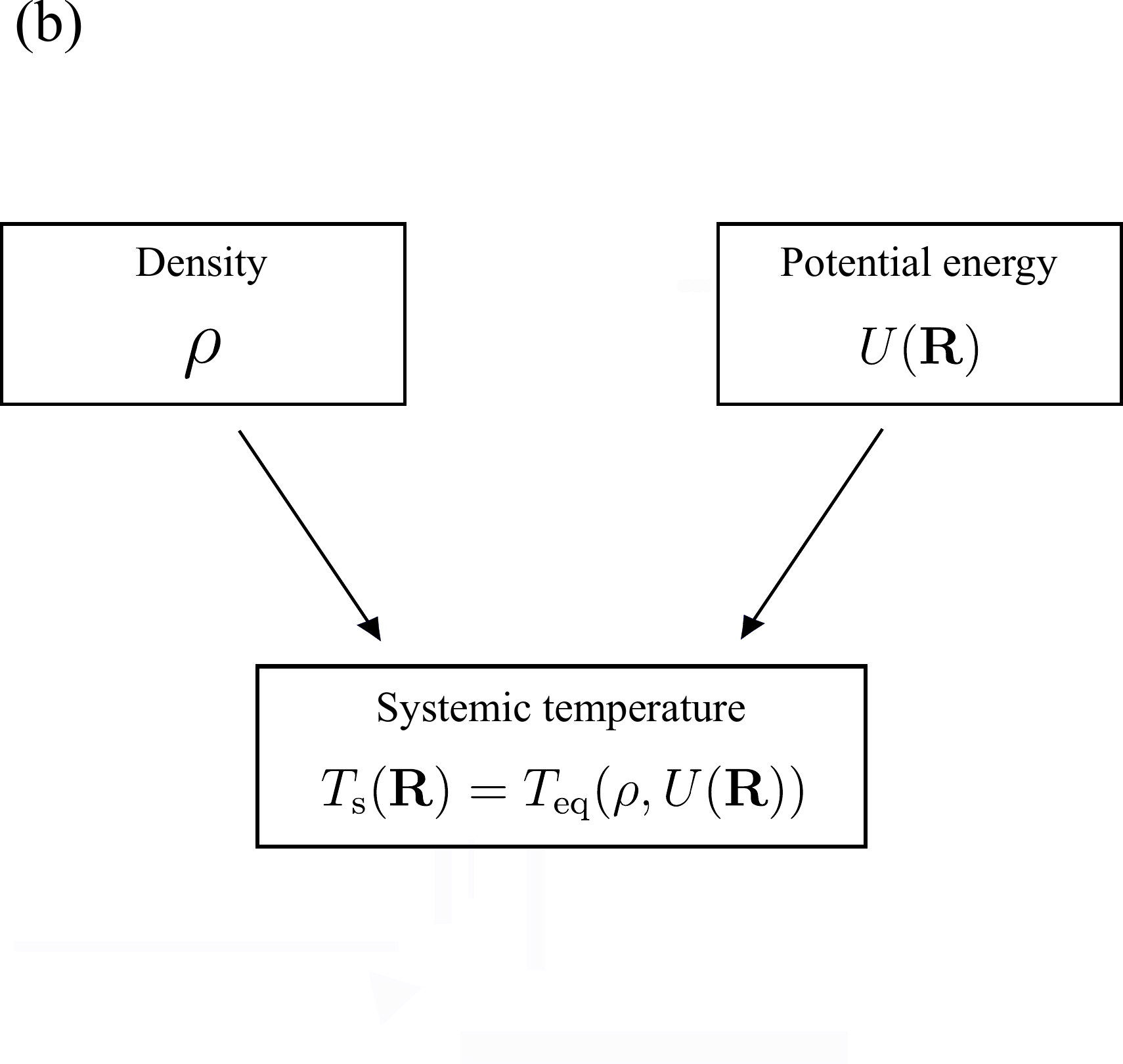}
	\end{center}
	\caption{\label{fig1} The systemic-temperature $\Ts(\bR)$ is defined for any configuration $\bR$ of any system, i.e., also for systems that are not R-simple. This figure summarizes the general situation.
	(a) illustrates that the systemic temperature for a given configuration is defined from its density $\rho$ and excess entropy $\Sex(\bR)$ (for an R-simple system, the excess entropy depends only on the configuration's reduced coordinates and $\Sex(\bR)$ may be replaced by $\Sex(\tbR)$). The systemic temperature is the temperature of the equilibrium state point with density $\rho$ and excess entropy $\Sex(\bR)$. In equilibrium at temperature $T$, the systemic temperature obeys $\Ts(\bR)\cong T$ with fluctuations that go to zero in the thermodynamic limit.	
	(b) shows how to identify $\Ts(\bR)$ in practice from the density and the potential energy: $\Ts(\bR)$ is the temperature of the equilibrium state point with density $\rho$ and average potential energy equal to $U(\bR)$.}
\end{figure*}

\section{Generalizing isomorph theory to systems that are not in thermal equilibrium}\label{general}

In this section we introduce systemic isomorphs as lines of constant excess entropy in the phase diagram defined by density and systemic temperature. Any configuration of an R-simple system identifies a systemic isomorph, whether or not the configuration is typical for an equilibrium state point.

\subsection{The systemic temperature $\Ts$}

In the expression for $\tbF$ in \eq{tbFeq}, the derivative of the thermodynamic equilibrium function $U(\rho,\Sex)$ with respect to $\Sex$ evaluated at $\Sex=\Sex(\tbR)$ appears. In thermal equilibrium this derivative is $T$, but in more general contexts a separate name is needed for it. For any configuration $\bR$ the \textit{systemic temperature} $\Ts(\bR)$ is defined \cite{dyr18} by

\be\label{Ts_def}
\Ts(\bR)
\,\equiv\,\left(\frac{\partial U(\rho,\Sex(\bR))}{\partial\Sex}\right)_\rho\,.
\ee
Just as the definition of $\Sex(\bR)$ in \eq{sex_def} does not assume hidden scale invariance, the same is the case for \eq{Ts_def}. We emphasize that it is always the equilibrium function $U(\rho,\Sex)$ that appears in \eq{Ts_def}. Thermal equilibrium is characterized by 

\be\label{eq_cond}
\Ts(\bR)\,\cong\,T\,,
\ee
in which the symbol $\cong$ indicates the existence of fluctuations that vanish in the thermodynamic limit.

Although \eq{Ts_def} may appear abstract, calculating $\Ts(\bR)$ in a simulation is straightforward. One makes use of the fact that $\Ts(\bR)$ is the equilibrium temperature $\Teq$ of the thermodynamic state point with the density of $\bR$ and with excess entropy equal to $\Sex(\bR)$. By the definition of $\Sex(\bR)$, this means that $\Ts(\bR)$ is the equilibrium temperature of the state point with density $\rho$ and average potential energy $U(\bR)$. Restricting henceforth to R-simple systems and using \eq{Sex_rc}, we summarize these identities as follows

\be\label{Ts_id}
\Ts(\bR)
\,=\,\Teq(\rho,\Sex(\tbR))
\,=\,\Teq(\rho,U(\bR))\,.
\ee
The last equality sign makes it possible to find $\Ts(\bR)$ from simulations by mapping out numerically the thermodynamic equilibrium function $U(\rho,T)$ and inverting it to obtain $\Teq(\rho,U)$. \Fig{fig1} illustrates the situation.

Note, incidentally, that when a configuration is scaled uniformly, $\Ts(\lambda\bR)$ is controlled by the equilibrium temperature's density dependence at fixed excess entropy,

\be\label{Ts_id2}
\Ts(\lambda\bR)
\,=\,\Teq(\lambda^{-3}\rho,\Sex(\tbR))\,.
\ee

\begin{figure*}[htbp!]
	\begin{center}
		\includegraphics[width=90mm]{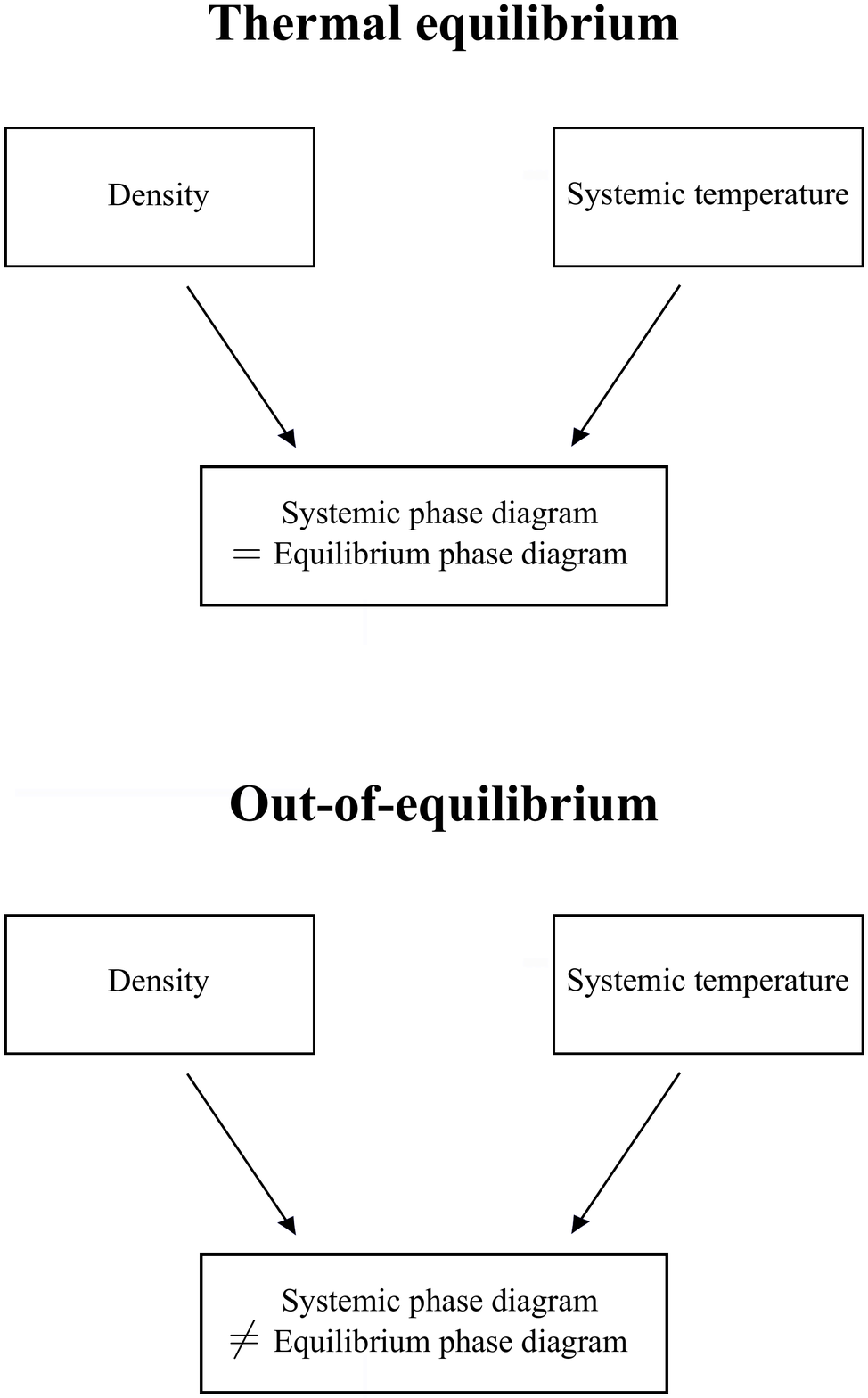}
	\end{center}
	\caption{\label{fig2} Relation between the systemic phase diagram defined by $\rho$ and $\Ts$ and the standard thermodynamic phase diagram defined by $\rho$ and $T$. In thermal equilibrium, $\Ts=T$ and the two phase diagrams are identical. In this case, systemic isomorphs reduce to equilibrium isomorphs \cite{IV,sch14}. Out-of-equilibrium situations are characterized by $\Ts\neq T$; here both phase diagrams are needed.}
\end{figure*}

\subsection{Systemic isomorphs}\label{si}

Any configuration $\bR$ is associated with a density $\rho$ and a systemic temperature $\Ts(\bR)$. Consequently, it may be mapped onto the two-dimensional ``systemic'' phase diagram defined by $\rho$ and $\Ts$. This is in contrast to the standard thermodynamic $(\rho,T)$ phase diagram onto which only equilibrium configurations may be mapped. 

\Eq{Ts_id} implies that each state point in the systemic phase diagram has a well-defined excess entropy, which is the excess entropy of the equilibrium state point with density $\rho$ and temperature equal to $\Ts$. Standard isomorphs are curves of constant excess entropy in the equilibrium thermodynamic phase diagram \cite{IV,sch14}. We define a \textit{systemic isomorph} as a curve of constant excess entropy in the systemic phase diagram. Since $\Sex$ at the systemic state point $(\rho,\Ts)$ is equal to the equilibrium excess entropy of the state point with density $\rho$ and temperature equal to $\Ts$ (\eq{Ts_id}), drawing the systemic isomorphs in the systemic phase diagram results in the very same set of curves as drawing the standard isomorphs in the thermodynamic phase diagram. The difference is that, as mentioned, any configuration is associated with a systemic isomorph whereas standard isomorphs involve only equilibrated configurations. The relation between the two phase diagrams is illustrated in \fig{fig2}.

\subsection{$\Ts/T$ controls the reduced-unit dynamics}\label{control}

This section establishes the condition for invariant dynamics along a systemic isomorph. The setting is that of an ensemble ${\mathfrak M}=\{\bR\}$ of generally non-equilibrium configurations $\bR$ of an R-simple system with same density and excess entropy. More precisely, it is assumed that the relative fluctuations of $\Sex(\bR)$ go to zero in the thermodynamic limit. This is the case if the mean-square potential-energy fluctuation is proportional to the system size, which applies for all systems without long-range interactions. Since $\Sex(\bR)$ depends only on the reduced coordinates of $\bR$ (\eq{Sex_rc}), scaling the configurations of ${\mathfrak M}$ uniformly to a different density moves ${\mathfrak M}$ along a systemic isomorph. The question is whether the dynamics is invariant if the temperature $T$ is adjusted appropriately in the process; the answer is yes as we shall see.

We regard both the density $\rho$ and the heat-bath temperature $T$ as externally controlled thermodynamic variables. The two standard realizations of this are Brownian (Langevin) dynamics and Nose-Hoover $NVT$ dynamics, each of which is considered below where a Gaussian isokinetic thermostat is also discussed.

Consider first Brownian dynamics, which was dealt with in detail in Ref. \onlinecite{dyr18} that introduced the concept of a systemic temperature in connection with physical aging. The Langevin equation of motion is \cite{cha43,reichl} 

\be\label{Lang1}
\dot\bR
\,=\,-\mu\nabla U(\bR)+\bet(t)\,.
\ee
Here $\mu$ is a constant and the noise vector $\bet(t)$ consists of Gaussian random variables $\eta_i(t)$ obeying

\be\label{stoj}
\langle \eta_i(t)\eta_j(t')\rangle
\,=\,
2\mu\,k_BT\, \delta_{ij}\delta(t-t')\,.
\ee
The corresponding Smoluchowski equation for the probability distribution $P(\bR,t)$ is %\cite{cha43,reichl}

\be\label{smol1}
\frac{\partial P(\bR,t)}{\partial t}
\,= \,\mu\,\nabla\cdot \Big(\nabla U(\bR)P(\bR,t)+k_BT\nabla P(\bR,t)\Big)\,,
\ee
which in reduced coordinates becomes \cite{dyr18}

\be\label{smol2a}
\frac{\partial P(\tbR,\tt)}{\partial\tt}
\,=\,\tilde{\nabla}\cdot\left(\frac{\Ts(\bR)}{T}\,\tilde{\nabla}\tSex(\tbR) P(\tbR,\tt)\,+\,\tilde{\nabla}P(\tbR,\tt)\right)\,.
\ee
Here one may replace $\Ts(\bR)$ by a constant $\Ts$ because the systemic temperature fluctuations as mentioned go to zero in the thermodynamic limit. \Eq{smol2a} has no reference to the density except via a possible density dependence of $\Ts$. This means that systems scaled to a different density will follow the same reduced-time evolution if $\Ts/T$ is the same. The condition for invariant dynamics along a systemic isomorph is therefore

\be\label{inv_cond}
\frac{\Ts}{T}
\,=\,{\rm Inv.}
\ee

We proceed to show that the same invariance condition applies for Nose-Hoover $NVT$ dynamics. If $\br_i$ and $\bp_i$ are, respectively, the position and momentum of particle $i$ and $Q$ is the (extensive) Nose-Hoover thermostat time  constant, the $NVT$ equations of motion \cite{frenkel} are

\begin{eqnarray}\label{NVT}
\dot{\br}_i\,&=&\,\frac{\bp_i}{m_i}\nonumber\\
\dot{\bp}_i\,&=&\,\bF_i\,-\xi{\bp_i}\\
\dot\xi\,&=&\,\left(\sum_i\frac{\bp^2_i}{2m_i}-\frac{3}{2}Nk_BT\right)/Q\nonumber\,.
\end{eqnarray}
These equations are made dimensionless by multiplication by combinations of the units given in \eq{red_units}:

\begin{eqnarray}\label{NVT2}
\frac{t_0}{l_0}\dot{\br}_i\,&=&\,\frac{t_0}{l_0}\frac{\bp_i}{m_i}\nonumber\\
\frac{t_0^2}{\avm l_0}\dot{\bp}_i\,&=&\,\frac{t_0^2}{\avm l_0}\bF_i\,-\,\frac{t_0^2}{\avm l_0}\xi{\bp_i}\\
t_0^2\dot\xi\,&=&\,t_0^2e_0\left(\sum_i\frac{\bp^2_i}{2e_0m_i}-\frac{3}{2}Nk_BT/e_0\right)/Q\nonumber\,.
\end{eqnarray}
The relevant reduced quantities are

\be
\tt \equiv t/t_0\,,
\,\tbr\equiv \br/l_0\,,
\,\tbp_i\equiv t_0\bp_i/(\avm l_0)\,,
\,\tbF_i\equiv l_0\bF_i/e_0\,,
\,\txi\equiv t_0\xi\,,
\,\tm_i\equiv m_i/\avm\,,
\,\tilde{Q}\equiv  Q/(e_0t_0^2)\,.
\ee
If a dot in connection with a reduced variable signals the derivative with respect to the reduced time $\tt$, the reduced $NVT$ equations of motion are 

\begin{eqnarray}\label{NVT3}
\dot{\tbr}_i\,&=&\,\frac{\tbp_i}{\tm_i}\nonumber\\
\dot{\tbp}_i\,&=&\,\tbF_i\,-\txi{\tbp_i}\\
\dot\txi\,&=&\,\left(\sum_i\frac{\tbp^2_i}{2\tm_i}-\frac{3}{2}N\right)/\tilde{Q}\nonumber\,.
\end{eqnarray}
These equations are independent of the density if the reduced force is a function of the reduced coordinates and if $\tilde{Q}$ is constant, i.e., $Q\propto\rho^{-2/3}$. The latter condition is not considered further because physically relevant quantities are insensitive to the precise value of $Q$. 

From \eq{tbFeq} and the defintion of $\Ts(\bR)$ (\eq{Ts_def}) the reduced collective force vector $\tbF$ is given by 

\be\label{tbFeq_2}
\tbF
\,=\,-\frac{\Ts(\bR)}{T}\,\tnabla\tSex(\tbR)\,.
\ee
Since the ensemble of states $\mathfrak{M}$ has non-extensive systemic temperature fluctuations, $\Ts(\bR)$ may be regarded as constant and \eq{tbFeq_2} becomes 

\be\label{tbFeq_3}
\tbF
\,=\,-\frac{\Ts}{T}\,\tnabla\tSex(\tbR)\,.
\ee
\Eq{tbFeq_3} implies that the reduced Nose-Hoover $NVT$ dynamics is invariant if temperature and systemic temperature along a systemic isomorph varies with density such that their ratio is constant (\eq{inv_cond}). 

Consider finally the Gaussian isokinetic thermostat, which in contrast to the Nose-Hoover algorithm keeps the kinetic energy strictly constant. The equations of motion \cite{tod07} are

\begin{eqnarray}\label{Gisokin}
\dot{\br}_i\,&=&\,\frac{\bp_i}{m_i}\nonumber\\
\dot{\bp}_i\,&=&\,\bF_i\,-\left(\frac{\sum_j\frac{\bp_j}{m_j}\cdot\bF_j}{\sum_j\frac{\bp^2_j}{m_j}}\right)\,{\bp_i}
\,.
\end{eqnarray}
The corresponding reduced equations are

\begin{eqnarray}\label{Gisokin2}
\dot{\tbr}_i\,&=&\,\frac{\tbp_i}{\tm_i}\nonumber\\
\dot{\tbp}_i\,&=&\,\tbF_i\,-\left(\frac{\sum_j\frac{\tbp_j}{\tm_j}\cdot\tbF_j}{\sum_j\frac{\tbp^2_j}{\tm_j}}\right)
\,{\tbp_i}\,.
\end{eqnarray}
Again, substituting \eq{tbFeq_3} into the above we see that these equations are invariant along a systemic isomorph if \eq{inv_cond} applies.

In all the cases discussed above, the reduced force is a function of the reduced coordinates times the systemic temperature divided by the bath temperature (\eq{tbFeq_3}). This is why there is invariance of the reduced dynamics along the systemic isomorphs when \eq{inv_cond} applies. Note that the invariant ratio $\Ts/T$ does not have to be constant in time. Note also that \eq{inv_cond} includes the thermal-equilibrium case of isomorph invariance: in equilibrium the systemic phase diagram reduces to the standard thermodynamic phase diagram with identical isomorphs and the ratio $\Ts/T$ is unity, i.e., the equilibrium dynamics is isomorph invariant.

\begin{figure*}[htbp!]
	\begin{center}
		\includegraphics[width=100mm]{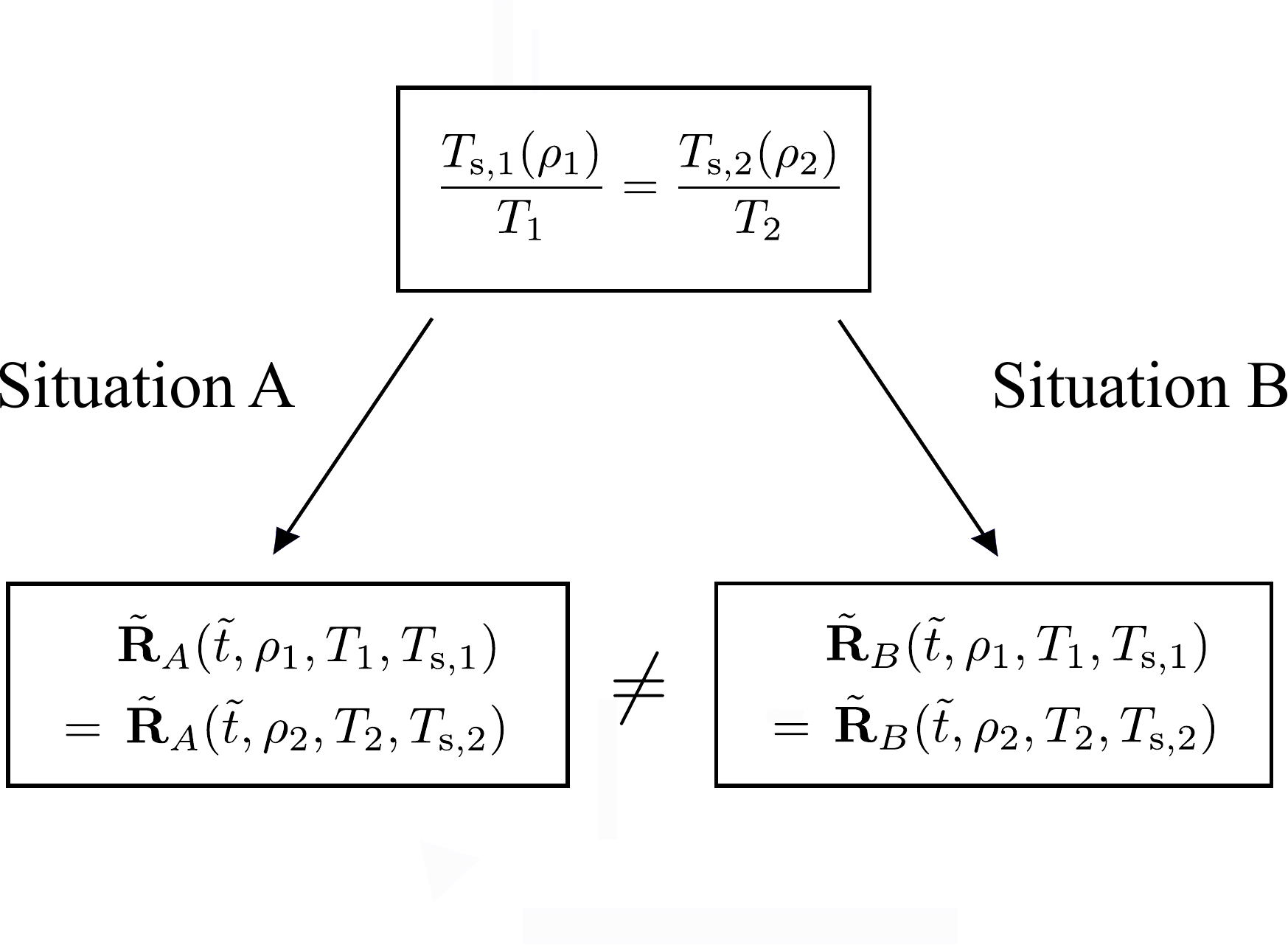}
	\end{center}
	\caption{\label{fig3}Knowledge of density, temperature, and systemic temperature is generally not enough to determine the physics. The figure illustrates this by considering two different non-equilibrium situations of the same system, A and B, at two densities with identical temperatures and systemic temperatures at both densities. For instance, this could correspond to a thermal history ending in a state with the same density, temperature, and potential energy, as that of an imposed flow. All non-equilibrium situations have the same systemic isomorphs, and if \eq{inv_cond} is satisfied, each non-equilibrium situation has invariant reduced-unit dynamics along the systemic isomorphs. The dynamics of situations A and B may well differ, however.}
\end{figure*}

We emphasize that the non-equilibrium system is \text{not} mapped to an equilibrium system at temperature $\Ts$ in the sense that all non-equilibrium physical quantities are the same as at the $T=\Ts$ equilibrium state point. By the definition of the systemic temperature, of course, the potential energy of the non-equilibrium system is that of the equilibrium state point with $T=\Ts$, but this property does not necessarily carry over to other quantities. In particular, two different non-equilibrium situations of the same system with same density, bath temperature, and systemic temperature, may have different properties and different dynamics. This is illustrated in \fig{fig3}.

\section{Questions and answers}

This section addresses three questions that may arise in view of the formalism developed above.

\subsection{How does the systemic temperature relate to other non-equilibrium temperatures?}

While the heat-bath temperature $T$ refers exclusively to the momentum degrees of freedom, a number of temperatures have been defined for non-equilibrium systems \cite{pug17}, which like $\Ts(\bR)$ depend only on the configurational degrees of freedom $\bR$ and reduce to $T$ in thermal equilibrium. This section discusses the relation between three such temperature and $\Ts(\bR)$. 

By being a function of $U(\bR)$, $\Ts(\bR)$ is a global rather than a local temperature, hence the name ``systemic''. This is in contrast to the configurational temperature defined \cite{pow05} by $k_B\Tcon(\bR)\equiv(\nabla U(\bR))^2/\nabla^2 U(\bR)$. $\Tcon(\bR)$ reflects how the potential energy varies close to $\bR$, whereas $\Ts(\bR)$ is determined by the potential energy $U(\bR)$ (\eq{Ts_id}). Clearly, these two temperatures cannot be identical in general. Interestingly, for R-simple systems there is a link between them. It is straightforward to show from \eq{fundeq} that $\Tcon(\bR)$ obeys 

\be\label{Tconsyst}
\frac{\Ts(\bR)}{\Tcon(\bR)}
\,\cong\,\frac{\tnabla^2\tSex(\tbR)}{(\tnabla\tSex(\tbR))^2}\,
\ee
in which $\cong$ signals deviations that go to zero in the thermodynamic limit. The right-hand side is isomorph invariant. This means that by \eq{inv_cond}, adjusting the heat-bath temperature $T$ with density along a systemic isomorph such that $T\propto\Tcon$ will lead to invariant dynamics. Note that that for equilibrium configurations \eq{Tconsyst} implies $\tnabla^2\tSex(\tbR)\cong(\tnabla\tSex(\tbR))^2$.

A glass is characterized by the so-called effective temperature $\Tef$ that quantifies the violation of the fluctuation-dissipation theorem (FDT) at long times \cite{bar00,leu09,cug11,pug17}. There is no FDT violation above the glass transition temperature $\Tg$, while below $\Tg$ the effective temperature reflects the frozen structure and \cite{leu09,cug11,pug17} 

\be\label{conj}
\Tef\,\simeq\,\Tg\,.
\ee
The systemic temperature behaves differently. Above $\Tg$ there is equilibrium and one has $\Ts=T=\Tef$, of course (we here and henceforth ignore that a glass usually forms from a supercooled liquid, a state that is not in true thermodynamic equilibrium but in a metastable equilibrium). Cooling below $\Tg$, however, the systemic temperature decreases continuously with $T$ also in the glass phase due to the decreasing potential energy of the vibrational degrees of freedom ($\Ts$ will be larger than $T$ due to the higher potential energy of the glass than of the metastable equilibrium liquid). Thus only close to $\Tg$ does one expect $\Ts\simeq\Tef\simeq\Tg$.

The effective temperature has been related to the thermodynamics \cite{leu09,cug11,pug17}. A possible link to the systemic temperature is that 

\be\label{conj2}
\Tef\,=\,\Ts\,,
\ee
at least in some situations. As argued above, this cannot apply for a glass because its systemic temperature is generally significantly lower than $\Tg$, but \eq{conj2} can possibly be obeyed in steady-state situations. In this connection we note that a two-temperature description of nonlinear rheology based on $\Tef$ and $T$ was proposed already twenty years ago \cite{ber00}.

In relation to viscous liquids and the glass transition, the so-called fictive temperature $\Tf$ is often used for interpreting experiments monitoring physical aging \cite{scherer}. The idea is that the structure of a glass is equal to that of the equilibrium metastable liquid at the temperature $\Tf$. Thus like the effective temperature (\eq{conj}), the fictive temperature of a glass is close to the glass transition temperature. Since this is not the case for the systemic temperature, which does not freeze upon cooling through the glass transition, we conclude that $\Tf\neq\Ts$.

My colleague Kristine Niss has recently proposed that $\Tf = \Tef$ \cite{nis17}. Niss has furthermore suggested that any state of a physically aging system can be mapped onto the equilibrium phase diagram and that this diagram must have lines of invariant structure. Although this differs from the above discussed mapping onto the systemic phase diagram, the two approaches are clearly closely related in view of the fact that in the present case curves of invariant structure and dynamics exist, which are identical in the systemic and the ``real'' phase diagrams.

\subsection{How to identify the systemic isomorphs in a computer simulation?}

Since a systemic isomorph is the same curve in the $(\rho,\Ts)$ phase diagram as a standard isomorph in the $(\rho,T)$ phase diagram, any method for generating the latter may be used for identifying the systemic isomorphs. A straightforward method integrates \eq{gamma} step-by-step by imposing density changes of typically a few percent, at each temperature recalculating the canonical averages in \eq{gamma} from a thermal equilibrium ($NVT$) simulation. Another general method is the ``direct isomorph check'' \cite{IV}. Here one uniformly scales equilibrium configurations obtained at one density, $\rho_1$, to a different density, $\rho_2$. According to \eq{isom}, the slope of a scatter plot of the potential energies of scaled versus unscaled configurations, i.e., of $U(\bR_2)$ versus $U(\bR_1)$ in which $\rho_1^{1/3}\bR_1=\rho_2^{1/3}\bR_2$, determines the temperature ratio $T_2/T_1$ for which $(\rho_1,T_1)$ and $(\rho_2,T_2)$ are on the same isomorph.

In the simplest approximation, the equilibrium isomorphs are given \cite{ing12a,boh12} by

\be\label{hrho_cond}
\frac{h(\rho)}{T}
\,=\,\rm{Const.}
\ee
in which the function $h(\rho)$ is defined as the quantity of dimension energy in the approximate equation $U(\bR)=h(\rho)\tilde{\Phi}(\tbR)+g(\rho)$ \cite{ing12,dyr13a}. For the LJ system, $h(\rho)$ is proportional to $(\gamma_0/2-1)(\rho/\rho_0)^4-(\gamma_0/2-2)(\rho/\rho_0)^2$ \cite{boh12} in which $\gamma_0$ is the density-scaling exponent at a reference state point of density $\rho_0$, a quantity that may be calculated from equilibrium fluctuations by means of \eq{gamma}. The corresponding systemic isomorphs are given by 

\be\label{hrho_cond2}
\frac{h(\rho)}{\Ts}
\,=\,\rm{Const.}
\ee

In many cases \eq{hrho_cond} gives a good representation of the equilibrium isomorphs, but for certain systems the more general equation $\Sex(\rho,T)=$ Const. must be used. This is the case when the density-scaling exponent $\gamma$ of \eq{gamma} is not only a function of density as implied by \eq{hrho_cond} \cite{ing12a}, which for instance applies in the gas phase of the EXP system \cite{EXPII} or at high temperatures for the LJ system \cite{V}. In this more general case, the invariance condition \eq{inv_cond} can still be fulfilled by a suitable choice of $T_2$. Suppose one studies an out-of-equilibrium system at density $\rho_1$ with temperature $T_1$, systemic temperature $\Tset$, and excess entropy $\Sex$. Then $\Tset=\Teq(\rho_1,\Sex)$ by \eq{Ts_id}. Being interested in the physics of the non-equilibrium system at density $\rho_2$, we ask whether a temperature $T_2$ exists resulting in invariant dynamics? The answer is yes because the following choice of $T_2$ does the job: 

\be\label{gen_cond}
T_2
\,=\,T_1\,\frac{\Tsto}{\Tset}
\,=\,T_1\,\frac{\Teq(\rho_2,\Sex)}{\Teq(\rho_1,\Sex)}\,.
\ee
In order to relate to previous works we do not refer below to \eq{gen_cond}, however, but to the simpler case \eq{hrho_cond} and \eq{hrho_cond2}. 

The above methods all involve performing equilibrium simulations. In steady-state situations it is possible to identify the systemic isomorphs directly from a non-equilibrium simulation. Consider two state points on a systemic isomorph with density $\rho_1$ and $\rho_2$. Non-equilibrium configurations with identical reduced coordinates are denoted by $\bR_1$ and $\bR_2$, and the time-averaged potential energies at the two densities are denoted by $U_1$ and $U_2$. \Eq{isom} was arrived at by Taylor expanding the basic relation \eq{fundeq}, and the same expansion may be carried out for a non-equilibrium system. The only difference is that the temperatures in \eq{isom} are replaced by systemic temperatures, i.e.,

\be\label{isom2}
\frac{U(\bR_1)-U_1}{k_B \Tset}\,\cong\,\frac{U(\bR_2)-U_2}{k_B\Tsto}\,.
\ee
It follows that the quantity $\Tsto/\Tset$ is the slope of a scatter plot of $U(\bR_2)$ versus $U(\bR_1)$, from which $\Tsto$ can be determined if $\Tset$ is known.

\subsection{What is the relation between systemic and equilibrium isomorphs?}

The systemic isomorphs are the same curves in the $(\rho,\Ts)$ phase diagram as the equilibrium isomorphs in the standard $(\rho,T)$ phase diagram. In view of this, one might be inclined to think that the process of going out of equilibrium simply corresponds to moving from an equilibrium isomorph to a different equilibrium isomorph. If this were a generally correct way of thinking about things, however, any non-equilibrium average should be equal to the corresponding equilibrium average evaluated at the temperature $\Ts$. While this may apply in some situations, as mentioned it cannot be general (\fig{fig3}). The non-equilibrium dynamics may drive the system to states that are unlikely at any temperature, for instance by breaking a spatial symmetry. This means that systemic isomorphs cannot be identified with equilibrium isomorphs. We need both phase diagrams.

In summary, even though the systemic isomorphs are the same curves in the $(\rho,\Ts)$ phase diagram for all non-equilibrium situations, the theory does \textit{not} imply identical physics for non-equilibrium situations with same density, temperature, and systemic temperature (\fig{fig3}). The \textit{only} prediction is that for each separate non-equilibrium situation, whenever \eq{inv_cond} applies, the reduced-unit structure and dynamics is invariant along the systemic isomorph in question.

\section{Examples}\label{app}

This section applies the systemic-isomorph concept to isomorph invariances identified in computer simulations of three different non-equilibrium systems. These were reported in previous \textit{Glass and Time} publications without consistent justifications.

\subsection{Steady-state Couette shear flows simulated by the SLLOD equations of motion [Separdar \textit{et al.}, J. Chem. Phys. \textbf{138}, 154505 (2013)]}\label{Leila}

An externally imposed steady-state shear flow drives a liquid away from equilibrium when the shear rate is large enough for the viscosity to become shear-rate dependent. Reference \onlinecite{sep13} studied nonlinear Couette shear flows of the standard single-component LJ system, as well as of the Kob-Andersen binary LJ mixture \cite{kob95} that is easily supercooled and brought into a highly viscous state. The systems were simulated by the SLLOD equations of motion \cite{eva84,tod07}, which utilize a Gaussian isokinetic thermostat. For both systems it was found that along standard equilibrium isomorphs:

\begin{enumerate}

\item For a given value of the reduced shear rate, the reduced radial distribution function is invariant.

\item For a given value of the reduced shear rate, the reduced transverse intermediate incoherent scattering function as a function of reduced time is invariant.

\item The reduced viscosity as a function of the reduced shear rate is invariant.

\item The reduced strain-rate-dependent parts of the potential energy is invariant as a function of the reduced shear rate.

\item The reduced strain-rate-dependent parts of the pressure is invariant as a function of the reduced shear rate.

\item The reduced strain-rate-dependent parts of the normal stress differences is invariant as a function of the reduced shear rate.

\end{enumerate}

In Ref. \onlinecite{sep13} these findings were rationalized by reference to the following equation

\be\label{iso_id}
U(\bR)
\,=\,k_BTf_I(\tbR)+g(Q)
\ee
in which the state point in question is denoted by $Q$ and $f_I(\tbR)$ is a function that may depend on the isomorph in question $I$. \Eq{iso_id} follows from the 2009 definition of isomorphs \eq{isomeq} that refers to thermal-equilibrium conditions \cite{IV}. Despite the fact that both systems of Ref. \onlinecite{sep13} were driven away from equilibrium as evidenced by the radial distribution functions changing significantly, \eq{iso_id} was used without further justification. In order to derive points 4-6 it was further assumed \textit{ad hoc} that $g(Q)$ is independent of the shear rate. In Ref. \onlinecite{sep13} isomorphs were defined as lines in the three-dimensional phase diagram defined by density, temperature, and shear rate. These 3d isomorphs turned out to ``project'' onto the equilibrium isomorphs of the $(\rho,T)$ equilibrium phase diagram. No explanation was offered of this observation, however, which is now seen to be a consequence of the definition and properties of systemic isomorphs (\sect{si}).

The justifications of the above invariances provided in ref. \onlinecite{sep13} are not satisfactory because they are based on equilibrium identities. How to explain the findings properly? For an R-simple system, the SLLOD equations of motion are isomorph invariant in reduced units provided $\Ts/T$ is the same along a given systemic isomorph. This is easy to prove by writing the SLLOD equations in reduced units and substituting \eq{tbFeq_3} into these. Suppose two state points $(\rho_1,T_1)$ and $(\rho_2,T_2)$ are on the same equilibrium isomorph. Then the following applies (compare \eq{hrho_cond})

\be\label{sseq2}
\frac{h(\rho_1)}{T_1}
\,=\,\frac{h(\rho_2)}{T_2}\,.
\ee
At the corresponding densities a systemic isomorph obeys \eq{hrho_cond2},

\be\label{sseq3}
\frac{h(\rho_1)}{\Tset}
\,=\,\frac{h(\rho_2)}{\Tsto}\,.
\ee
Dividing \eq{sseq2} by \eq{sseq3} leads to the required invariance condition \eq{inv_cond},

\be\label{sseq}
\frac{\Tset}{T_1}
\,=\,
\frac{\Tsto}{T_2}\,.
\ee
To be specific, consider a steady-state shear flow at density $\rho_1$ and temperature $T_1$ with the reduced-coordinate solution of the SLLOD equations of motion $\tbR_1(\tt)$. Because of \eq{sseq}, for the same reduced shear rate $\tbR_1(\tt)$ will also solve the reduced SLLOD equations of motion at density $\rho_2$ and temperature $T_2$. This established points 1-3 above without reference to the equilibrium identity \eq{iso_id}. Note that the condition of a constant $\Ts/T$ means that along any systemic isomorph, \eq{iso_id} can be rewritten as $U(\bR)=k_B\Ts F_I(\tbR)+g(Q)$, which may be derived by Taylor expanding \eq{fundeq} to first order in the excess entropy. In other words, \eq{iso_id} is actually correct although its justification was not. 

To derive point four above, if $U(\rho,\Sex)$ as previously is the equilibrium thermodynamic functions and $\tdg$ is the reduced shear rate, we make a first-order Taylor expansion of \eq{fundeq} in $\Sex$ around equilibrium ($\Ts=T$, $\tdg=0$). The steady-state flow average potential energy $U(\rho,T,\tdg)$ is, by definition of the nonequilibrium excess entropy $\Sex(\tdg)\equiv \Sex(\rho,\Ts)$, equal to $U(\rho,\Sex(\tdg))$, and we therefore have [identifying $U(\rho,T)$ with $U(\rho,\Sex)$]

\be\label{U_eq}
U(\rho,T,\tdg)
\,=\, U(\rho,\Sex(\tdg))
\,=\,U(\rho,T)+T (\Sex(\tdg)-\Sex) \,+\,...\,.
\ee
This implies 

\be\label{U_eq2}
\frac{U(\rho,T,\tdg)-U(\rho,T)}{k_BT}
\,\cong\, \tSex(\tdg)-\tSex\,.
\ee
The left-hand side is the reduced strain-rate-dependent part of the potential energy. The right-hand side is isomorph invariant for any given value of $\tdg$. This demonstrates point four above. The numerical data of Fig. 7 in Ref. \onlinecite{sep13} show a small, but systematically increasing deviation from isomorph invariance with increasing reduced shear rate; this is consistent with the fact that higher-order terms are ignored in \eq{U_eq}.

We next turn to point five, the isomorph invariance of the reduced pressure difference. The pressure $p$ is related to the virial $W$ by $pV=Nk_BT+W$. Thus the reduced pressure difference is given by $[p(\rho,T,\tdg)-p(\rho,T)]/(\rho k_BT)=[W(\rho,T,\tdg)-W(\rho,T)]/(N k_BT)$ in which $p(\rho,T)$ and $W(\rho,T)$ are the equilibrium pressure and virial. Because the microscopic virial is defined by $W(\bR)\equiv(\partial U(\bR)/\partial\ln\rho)_{\tbR}$ \cite{dyr18a}, \eq{fundeq} implies $W(\bR)=W(\rho,\Sex(\tbR))$ in which $W(\rho,\Sex)\equiv(\partial U/\partial\ln\rho)_{\Sex}$ is the thermodynamic equilibrium virial. For the averaged quantities this implies that $W(\rho,T,\tdg)-W(\rho,T)=W(\rho,\Sex(\rho,\Ts))-W(\rho,\Sex(\rho,T))=W(\rho,\Sex(\tdg))-W(\rho,\Sex)$. Taylor expanding this to first order and using the thermodynamic identities $W=(\partial U/\partial\ln\rho)_{\Sex}$ and $T=(\partial U/\partial\Sex)_{\rho}$ leads to $W(\rho,T,\tdg)-W(\rho,T)\cong(\partial T/\partial\ln\rho)_{\Sex} (\Sex(\tdg)-\Sex)$. By the definition of the density-scaling exponent $\gamma$ in \eq{gamma} this implies

\be\label{W_eq}
\frac{W(\rho,T,\tdg)-W(\rho,T)}{k_BT}
\,\cong\,\gamma\left(\tSex(\tdg)-\tSex\right)\,.
\ee
This proves the isomorph invariance of the reduced strain-rate-dependent part of the pressure for fixed $\tdg$. \Eq{U_eq2} and \eq{W_eq} imply that the reduced pressure differences equals $\gamma$ times the reduced potential-energy difference per particle. This is consistent with the numerical data of Ref. \onlinecite{sep13}. 

For point six above, note first that in terms of the stress tensor $\sigma_{\mu\nu}$, the normal pressure difference is $(\sigma_{xx}-\sigma_{yy})/2$ in which $x$ is the flow direction and $y$ is the direction the velocity gradient. The $xx$ stress tensor is given by the following sum over all particles, $\sigma_{xx}=(1/V)\sum_{ij}(x_i-x_j)F_x^{ij}$, in which $F_x^{ij}=-\partial U(\bR)/\partial(x_i-x_j)$, and a similar expression applies for $\sigma_{yy}$. In this way one relates to $\tbF(\tbR)$, and it is now easy to establish the required systemic isomorph invariance of the reduced normal stress for any given $\tdg$.

\subsection{Flow-event statistics for athermal plastic flows of glasses [Lerner \textit{et al.}, Phys. Rev. E \textbf{90}, 052304 (2014)]}\label{Edan}

Ref. \onlinecite{ler14} presented computer simulations of zero-temperature glasses subject to an imposed shear flow. Samples were prepared by a rapid quench from the liquid. At any given time there is mechanical equilibrium, i.e., the force on each particle is zero. A steady-state flow situation consists of a continuous increase of the stress with time as the strain increases, interrupted by discontinuous stress drops deriving from avalanches in the solid. The two main models considered were the Kob-Andersen binary LJ system and its repulsive version in which the $r^{-6}$ terms are positive instead of negative. The observables were the steady-state probability distributions of stress drops, potential-energy drops, and strain increases between two stress drops.

By scaling with the function $h(\rho)$ encountered above in connection with \eq{hrho_cond}, it was shown in Ref. \onlinecite{ler14} how the observables at different densities can be predicted from simulations at a single reference density. This was justified by dimensional analysis: at zero temperature the only quantity of dimension energy is the function $h(\rho)$ (compare Sec. IV B) \cite{dyr13a}. For each of the two systems studied, $h(\rho)$ was evaluated by computer simulations of the equilibrium liquid phase. 

How can one understand that the liquid's $h(\rho)$ controls the zero-temperature plastic flow physics? To answer this, note that the preparation of the $T=0$ amorphous solid by quenching a liquid at the reference density leads to a sample with $\Ts>0$. The precise value of $\Ts$ is not important; $\Ts$ is significantly below the glass transition temperature of the quench, $T_g$, because the vibrational degrees of freedom at $T_g$ still have a sizable potential energy. Changing the density of the zero-temperature glass by compressing or expanding the boundaries induces a virtually uniform scaling of all particle coordinates (this is a consequence of \eq{hsi} \cite{dyr18}). Consequently, by \eq{Sex_rc} glasses of different density obtained by scaling a reference-density glass will belong to the same systemic isomorph. The function $h(\rho)$ in \eq{hrho_cond2} should be calculated for the equilibrium crystalline phase if the glass potential energy is below that of the crystal at melting at the density in question. The difference between the liquid and crystal $h(\rho)$ functions at the same density is only minor, however \cite{ped16}. Thus the systemic isomorph identifies the energy scale to be used in predicting the probability distributions of flow-event characteristics at different densities from observations at the reference density -- the relevant energy scale is $h(\rho)$ or, equivalently, $k_B\Ts(\rho)$ (compare \eq{hrho_cond2} and the discussion below in \sect{energy_unit}).

\subsection{Sheared glassy systems [Jiang \textit{et al.}, Phys. Rev. E \textbf{100}, 053005 (2019)]}\label{Yonglun}

A comprehensive simulation study of sheared finite-temperature glasses was presented recently \cite{jia19}. This case is in-between the SLLOD-simulated steady-state Couette flow of liquids (\sect{Leila}) and zero-temperature amorphous-solid shear deformations (\sect{Edan}). Focusing on the Kob-Andersen binary LJ mixture,  Ref. \onlinecite{jia19} demonstrated invariance of the following quantities along a low- and a high-temperature isomorph in the glass: 

\begin{enumerate}
	\item The reduced radial distribution function.
	\item The reduced average flow stress and its standard deviation.
	\item The reduced stress autocorrelation function as a function of strain interval.
	\item Histograms of reduced stress changes over a given strain interval for given reduced shear rate.
	\item The Fisher-Pearson skewness of the reduced stress-change distributions as a function of strain interval for given reduced shear rate.
	\item The incoherent intermediate scattering function (transverse direction) as a function of the reduced time for a given reduced shear rate.
	\item The reduced mean-square displacement (transverse direction) as a function of the reduced time for a given reduced shear rate.
\end{enumerate}
These invariants were justified by reference to standard isomorph theory. Indeed, the two glass-state isomorphs were generated by numerically integrating \eq{gamma} ignoring the fact that a glass is an out-of-equilibrium state.

Given that isomorphs are defined by reference to thermal equilibrium, not to non-equilibrium states like a glass, the question is how to justify the findings in a consistent setting. The answer is that the glass isomorphs studied in Ref. \onlinecite{jia19} are, in fact, systemic isomorphs obeying the invariance condition \eq{inv_cond}. To see this, note that the isomorphs in Ref. \onlinecite{jia19} obey \eq{hrho_cond}, while the corresponding systemic isomorphs obey \eq{hrho_cond2}. As in \sect{Leila}, dividing these two identities by one another leads to $\Ts/T=$ Inv. along the two isomorphs.

\section{Some further connections}

This section discusses briefly connections to non-equilibrium situations different from flows.

\subsection{Granular media}\label{granular}

Granular media has been an important area of research for several years \cite{rad17,bau18,beh19}. In 1989 Edwards and coworkers introduced the \textit{compactivity} concept in a daring thermodynamic approach to the subject \cite{edw89,meh89}. The idea was that, despite the absence of anything like a dynamic equilibrium involving transitions between several states,  ``when $N$ grains occupy a volume $V$ they do so in such a way that all configurations are equally weighted'' \cite{bau18}. Volume here plays the role of energy in conventional statistical mechanics, and for each volume $V$ the logarithm of the number of states defines an entropy function, $S=S(V)$. The compactivity $X$ is then defined in analogy to temperature by

\be\label{comp_def}
X
\,\equiv\,\frac{d V}{d S}\,.
\ee
Ref. \onlinecite{edw89} noted that ``the volume therefore depends on the configuration of the particles -- unlike the conventional case where the volume is set externally, and only the energy depends on the configuration of the particles''. Thus, via its volume each configuration has an entropy. This is analogous to the microscopic excess entropy defined in \eq{sex_def}. Likewise, the compactivity is analogous to the systemic temperature. An important difference, though, is that only jammed configurations were considered by Edwards and coworkers whereas we allow for all possible configurations.

Despite some initial skepticism, the Edwards approach to granular media turned out to be very useful \cite{bau18}. This gives rise to optimism that the non-equilibrium isomorph formalism will also be useful.

\subsection{Physical aging}\label{aging}

A glass is produced by continuously cooling a liquid below its melting point until it falls out of metastable equilibrium and solidifies \cite{dyr06}. As pointed out by Simon almost hundred years ago \cite{sim31}, any glass approaches very slowly the metastable equilibrium supercooled liquid phase at the actual temperature. This process is referred to as physical aging \cite{scherer,str78,hod95,mck17,rut17}. In practice, physical aging of a glass prepared from the liquid by slow cooling can only be observed by careful long-time annealing experiments right below the glass transition temperature \cite{scherer,mck17,hec19}.

Based on a Brownian dynamics approach, Ref. \onlinecite{dyr18} showed that physical aging is controlled by $\Ts/T$ (\eq{inv_cond}); the same applies if Nose-Hoover dynamics is used (\sect{control}). Physical aging differs from the steady-state situations discussed in \sect{app} because in physical aging $\Ts$ changes continuously with time. In fact, $\Ts(t)\to T$ as $t\to\infty$ as the system eventually equilibrates at the ``annealing'' temperature $T$. In this case, the time evolution of $\Ts$ is itself determined by the aging process. Isomorph invariance is predicted for annealing at different densities: if the starting conditions have the same $\Sex$, i.e., are on the same systemic isomorph, and if the annealing temperatures refer to the same equilibrium isomorph, the aging processes are identical in reduced coordinates \cite{dyr18}.

\subsection{Active matter}\label{active}

An intriguing area of research is the dynamics of active matter like bacteria or colloids propelled by chemical reactions \cite{cat12,mar13a,mag15}. Active matter consists of particles that absorb energy from the environment and convert it into various kinds of persistent motions. This leads to several spectacular phenomena like a tendency for particles to accumulate at solid walls or the formation of bound states between purely repulsive objects. In contrast to the cases considered above, active matter breaks time-reversal invariance. 

A simple model is the ``run and tumble model'' in which there is persistent motion of particles over a certain time interval until they suddenly change to a random new direction \cite{cat12,mar13a}. This feature is captured qualitatively by adopting a standard Langevin equation with, however, colored noise instead of the usual white noise of Brownian dynamics \cite{fil12,mag15}. A systemic temperature may be introduced for this active-matter model if the potential-energy function has hidden scale invariance. It would be interesting to investigate isomorphs of such a non-time-reversal-invariant system and, possibly, to connect the systemic temperature to the effective temperature $\Tef$ of FD-theorem violations that has also been discussed in connection with active matter \cite{loi08}.

\section{What is the correct energy unit defining reduced quantities?}\label{energy_unit}

A reduced quantity is arrived at by making the quantity in question dimensionless by multiplication by a proper combination of the units of \eq{red_units}. The time unit is derived from the length and energy units, which are more fundamental in the present context.

Both in and out of equilibrium, the length unit is the average nearest-neighbor distance between particles. However, when the system is not in equilibrium, two possible temperatures may be used for defining the energy unit $e_0$: $T$ or $\Ts$. The heat-bath temperature $T$ refers to the momentum degrees of freedom while $\Ts$ refers to the configurational degrees of freedom. Since the latter are central in isomorph theory, an obvious question is whether one should use as energy unit the systemic temperature instead of the present $k_BT$, 

\be\label{alt_en_unit}
e_0
\,=k_B\Ts\,\,\,(?)
\ee
Doing so would provide a density-dependent energy unit, which can be used also for a $T=0$ glass. This would justify the use of the function $h(\rho)$ in Ref. \onlinecite{ler14} as the energy scale of the flow-property probability distributions for glasses (\sect{Edan}) because along a systemic isomorph one has $\Ts\propto h(\rho)$ according to \eq{hrho_cond2}. 

Reference \onlinecite{jia19} discussed the possibility of using $h(\rho)$ as energy unit instead of $k_BT$. It was noted that if this is done, the reduced quantities along the two isomorphs studied are much closer to each other than when using $k_BT$ as the energy unit. It was moreover pointed out that while $e_0=k_BT$ implies that the time unit reflects how long it takes for free thermal-velocity motion to cover the nearest-neighbor length $l_0$, using instead $e_0\propto h(\rho)$ corresponds better to the vibrational time scale of particles in a glass.

In equilibrium, $\Ts=T$ and the two possible energy units coincide. Interestingly, along any systemic isomorph with dynamic invariance, the condition $\Ts/T=$ Inv. implies that the reduced equations of motion are mathematically equivalent for the two possible choices of energy unit.

\section{Summary}

Isomorphs may be defined also for R-simple systems that are not in thermal equilibrium. ``Systemic'' isomorphs are curves of constant excess entropy just as the original thermal-equilibrium isomorphs, but located in the systemic phase diagram defined by density and systemic temperature. For equilibrium systems, the systemic phase diagram reduces to the standard density-temperature thermodynamic phase diagram and the systemic isomorphs reduce to equilibrium isomorphs. The condition for invariant dynamics along a systemic isomorph is $\Ts/T=$ Inv. The generalized isomorph theory rationalizes a number of previous findings that were at the time not explained within a consistent setting.

\newpage
\acknowledgments{In preparing this paper I have benefited greatly from discussions with Kristine Niss, Lorenzo Costigliola, Nick Bailey, Shibu Saw, and Thomas Schr{\o}der. This work was supported by the VILLUM Foundation's \textit{Matter} grant (No. 16515).}

\end{document}